\documentclass[twocolumn,showpacs,aps,pra,amsmath,amssymb]{revtex4}
\usepackage{graphicx}
\usepackage{dcolumn}
\usepackage{bm}
\usepackage{txfonts}
\usepackage{amsmath,epsfig}
\usepackage{epstopdf}
\newcommand{\ket}[1]{\left\vert#1\right\rangle}
\newcommand{\s}{\uparrow}
\newcommand{\g}{\downarrow}
\newcommand{\ug}{\!=\!}
\newcommand{\eq}{Eq.~}
\newcommand{\eqs}{Eqs.~}
\newcommand{\fig}{Fig.~}
\newcommand{\figs}{Figs.~}
\newcommand{\cfr} {cfr.~}
\newcommand{\ie} {i.e.~}
\newcommand{\eg} {e.g.~}
\newcommand{\secn}{Sec.~}
\newcommand{\secns}{Secs.~}

\newcommand{\bra}[1]{\left\langle#1\right\vert}
\newcommand{\tr}{{\rm Tr}}

\begin{document}
\author{Francesco Ciccarello\mbox{$^{1}$}}
\author{Sougato Bose\mbox{$^{2}$}}
\author{Michelangelo Zarcone\mbox{$^{1}$}}
\affiliation{ \mbox{$^{1}$} CNISM and Dipartimento di Fisica e
Tecnologie Relative, Universit\`{a} degli Studi di Palermo, Viale
delle
Scienze, Edificio 18, I-90128 Palermo, Italy \\
\mbox{${\ }^{2}$} Department of Physics and Astronomy,
University College London, Gower Street,
London WC1E 6BT, United Kingdom}

\begin{abstract}

We consider a one-dimensional (1D) structure where non-interacting spin-$s$ scattering centers, such as quantum impurities or multi-level atoms, are embedded at given positions. We show that the injection into the structure of unpolarized flying qubits, such as electrons or photons, along with {path} detection suffice to accomplish spin-state teleportation between two centers via a third ancillary one. {No action over the internal quantum state of both the spin-$s$ particles and the flying qubits is required. The protocol enables the transfer of quantum information between well-seperated static entities in nanostructures by exploiting a very low-control mechanism, namely scattering.}

 \end{abstract}


\title{Teleportation between distant qudits via scattering of mobile qubits}
\maketitle

\section{Introduction}

Teleportation of a quantum state between two remote parties \cite{bennett} is a striking prediction of Quantum Mechanics. While being a significant manifestation of the non-local correlations inherent in entangled states \cite{nielsen} in constrast to the argument by Einstein, Podolsky and Rosen \cite{epr} a number of experimental tests confirmed its effectiveness in various physical scenarios \cite{bowm-boschi}. Besides its intriguing features, quantum teleportation has pivotal applications in the accomplishment of Quantum Information Processing (QIP) tasks. Significantly, teleportation can be harnessed as a resource to perform universal quantum computation \cite{tbqc}. This model for computation is indeed a prominent example of the emerging paradigm of measurement-based quantum computation (MQC)  \cite{browne} and is the focus of ongoing experimental work \cite{pan-fiu}.
While most of the schemes to implement teleportation focused on Quantum Optics/Cavity Quantum Electrodynamics setups \cite{no-bell, telep-QED}, proposals in solid-state scenarios have been put forward more recently \cite{telep-ss}.

A topical issue in the current quest for reliable ways to process quantum information is the pressing need for schemes able to work in limited-control situations. For instance, a general feature of MQC models of quantum computation \cite{browne}, in particular those based on teleportation \cite{tbqc}, is the replacement of time-controlled gating \cite{nielsen} with simple local measurements in order to process quantum information. Recently, a number of schemes for the generation of entangled states based on scattering processes have been put forward \cite{imps,ciccarello,resil-refs,mappaP, habgood,mappaNP,yuasa,daniel}. Such methods fit quite naturally into the scenario of limited-control QIP. Scattering is indeed a typical process occurring under low-control conditions: two or more particles are prepared and collided and, once the scattering event has taken place, measurements are performed, hence without any direct access to the interaction process. In such proposals, the typical strategy to establish entanglement is to scatter flying mediators, such as electrons or photons, between remote non-interacting quantum scattering centers, such as magnetic impurities or artificial atoms. The centers' cross-talk mediated by the flying particles allows to distribute entanglement between them. Suitable post-selection of the mediators' internal state, working conditions yielding additional symmetries and the use of a stream of scattering particles (rather than single ones) typically optimize entanglement generation. Major advantages of these strategies are the remarkable resilience against static disorder and non-optimal parameters \cite{resil-refs,mappaP}, decoherence affecting the centers \cite{mappaNP} and, as anticipated, the only mild control required over interaction times, which is inherited from the very nature of scattering processes. Very recently, it has been demonstrated \cite{mappaNP,yuasa} that efficient entanglement generation can be achieved even by fully relaxing any preparation and post-selection of the mediators' internal quantum state: mere injection of a manifold of unpolarized mobile particles that have trespassed the scattering region, as recorded by a {Geiger-like detector \cite{geiger}, can suffice to distribute maximum entanglement \cite{mappaNP, yuasa}. Latest achievements have shown that such a strategy can be modified so as to deterministically establish maximum entanglement \cite{daniel}.}

So far, the work carried out  along this line has essentially focused on generation of entangled states \cite{imps,ciccarello,resil-refs,mappaP, habgood,mappaNP, yuasa,daniel}. A natural question to ask is whether the implementation of tasks useful for QIP other than mere preparation of non-classical states can benefit from the attractive advantages of such scattering-based strategies. One simplest task in this category, and one which is extremely useful in a QIP context, is to move quantum information encoded in static qubits from one place to another. A possibility of accomplishing this is through teleportation.   Indeed, while  
it is quite natural to expect that a stream of mobile particles scattering between distant centers correlate them so as to establish entanglement, the possibility that processes of this sort can carry out a genuine quantum {\it algorithm}, such as teleportation, does not appear a foregone conclusion. On the other hand, the ability to prepare and measure entangled states is known to allow for teleportation. In particular, although the measurement of maximally entangled states is in general quite a demanding task, it is now well-established that teleportation can be accomplished by getting around direct joint measurements  \cite{no-bell}. This typically occurs when ancillary degrees of freedom are added to the particles between which teleportation is to be performed and made interact with them. At the end of the interaction process,  one can map certain local measurements onto the ancillary systems into {\it effective}, and thus indirect, Bell-state measurements \cite{no-bell, nota-bell}.

In this paper, we show that teleportation of quantum states between non-interacting parties can be performed via a scattering-based strategy. In particular, we demonstrate that the injection of a manifold of scattering particles along with standard Geiger-like detection are enough in order to teleport an arbitrary quantum state between two remote scattering centers with the assistance of an ancillary center.
Our scheme in fact does not require to take any action over either the centers or the internal degrees of freedom of the flying particles. Hence, it fits particularly well into a scenario where only limited or no access to quantum registers is available. 

 The motivation behind our focusing on teleportation is twofold. First, as discussed, this phenomenon has several and important applications \cite{tbqc}. Second, it particularly well embodies a paradigmatic process where quantum information flows over space. One would not expect that such a task could be successfully implemented through scattering events, where a number of quantum degrees of freedom such as those of the involved scattering mediators enter the dynamics. The counter-intuitiveness of our scheme is further strengthened by the lack of any requirement over the internal quantum state of the mobile particles: as we show later, these are sent each in a maximally mixed state and without requiring any post-selection of their internal state once they are scattered off. More specifically, we use a suitable quantum map that describes how the quantum state of the centers is changed by their interaction with each scattering particle and illustrate how {\it repeated} applications of such map, corresponding to the passage of a {\it stream} of mediators, makes teleportation effective with significant probability. 
 
 Significantly, our scheme is free from any requirement over the Hibert space dimension of each scattering center. It therefore allows for teleportation of the quantum state of \emph{qudits} (\ie particles having a $d$-dimensional Hilbert space with $d\!\ge\!2$). To the best of our knowledge, this is the first proposal for teleportation in a solid-state setting which is not restricted only to transmission of qubits.

The present paper is structured as follows. In \secn  \ref{setup} we introduce the setup and Hamiltonian and briefly outline the approach we used to tackle the problem. In \secn  \ref{protocol}{, the central section of this work,} we present our teleportation protocol and illustrate its effectiveness through some plots. In \secn \ref{mechanism}, we discuss more in detail the features of our approach, introduce the aforementioned quantum map and demonstrate the working principle of the scheme. In \secn  \ref{resilience} we address some issues that may arise in an experimental implementation such as the resilience of the protocol performances against static disorder and imperfect setting of resonance conditions.{We also comment on the feasibility of one of the few requirements of the scheme, \ie the need for switching off some couplings at certain stages of the teleportation process.}

Finally, in \secn \ref{conclusions} we draw our conclusions. 

\section{Setup} \label{setup}

We consider a one-dimensional (1D) wire along the $x$-axis where spin-1/2 mobile particles $e$ can propagate. In practice, $e$ can be embodied by either a conduction electron traveling in a quasi-1D semiconductor nanowire or carbon nanotube (see \cite{single-electron}  and references therein) or a photon propagating along a 1D solid-state waveguide such as a GaAs/GaN nanowire sustaining two frequency-degenerate orthogonally polarized modes \cite{mappaP,mappaNP}. Three spin-$s$ quantum scattering centers, labeled 1, 2 and 3, lie at $x\ug x_1\ug 0$, $x\ug x_2\ug d_{12}$ and $x\ug x_3\ug d_{23}$, respectively [$d_{12}$ ($d_{23}$) is the distance between centers 1-2 (2-3)]. In the case of electrons, the centers can be embodied by quantum impurities such as magnetic impurities or quantum dots (QDs) \cite{ciccarello,mappaP,mappaNP}, whereas in the photonic implementation these can be realized through multi-level artificial atoms such as such as InAs/GaInN QDs \cite{mappaP, mappaNP} or nitrogen-vacancy centers in diamond ~\cite{photonmodel} (in the case that $s\ug 1/2$ these are three-level $\Lambda$-type atoms with a two-fold degenerate ground state and one virtually-excited state \cite{mappaP,mappaNP}). Finally, a Geiger-like detector $D$ located at one end of the setting fires whenever it records the presence of a mobile particle $e$ regardless of its spin state. The whole setup is sketched in \fig 1. 

Whenever a mobile particle $e$ is at $x\ug x_i$ ($i\ug 1,2,3$), its spin interacts with that of the corresponding center according to the Hamiltonian 
\begin{figure}
\centerline{\includegraphics[width=0.35\textwidth]{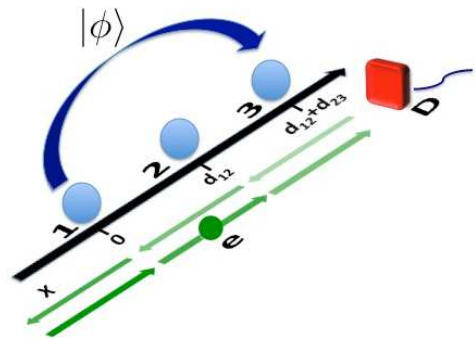}} 
\caption{(Color online) Proposed setup for the implementation of our scheme. A mobile quantum particle $e$ can propagate along the $x$-axis. Three spin-$s$ scattering centers , labeled 1, 2 and 3, lie at $x\ug x_1\ug 0$, $x\ug x_2\ug d_{12}$ and $x\ug x_3\ug d_{12}\!+\! d_{23}$, respectively. Whenever $e$ is at $x\ug x_i$ (${i=1,2,3}$), its spin interacts with that of the corresponding center $i$. The protocol allows to teleportat an unknown quantum state $\ket{\phi}$ between particle 1 (Alice) and particle 3 (Bob).\label{Fig1}}
\end{figure}
 \begin{equation} 
\label{H}
\hat{H}=\hat{H}_{kin}\!+\!\hat{V}
\end{equation}
with 
\begin{equation} 
\label{Vel}
\hat{V}\ug 
\sum_{i=1,2,3}J_i \,\hat{\mbox{\boldmath$\sigma$}}\cdot \hat{\mathbf{S}}_i\,\delta(x-x_i)\,\,,
\end{equation}
for the electronic setup and 
\begin{equation} 
\label{Vph}
\hat{V}\ug 
\sum_{i=1,2,3}J_i \,\left( \hat{\sigma}_x\hat{S}_{ix}+\hat{\sigma}_y\hat{S}_{iy} \right)\,\delta(x-x_i)\,\,,
\end{equation}
for the photonic setting.
{In \eqs (\ref{H})-(\ref{Vph}), $\hat{\mbox{\boldmath$\sigma$}}$ and $\hat{\mathbf{S}}_i$ are the (pseudo) spin operators of $e$ and the $i$-th center, respectively, $\hat{H}_{kin}$ is the kinetic Hamiltonian of $e$ and $J_i$'s are interaction strengths (in units of frequency times length). 
Throughout this paper we set $\hbar\ug1$. When $e$ is embodied by a photon its pseudo-spin $\hat{\mbox{\boldmath$\sigma$}}$ is that associated with the Hilbert space spanned by two orthogonal polarization states (for its rigorous definition see Ref.~\cite{mappaP}). Concerning the kinetic Hamiltonian, for the implementation using electrons we have $H_{kin}\ug\hat{p}^2/(2m^*)$ ($\hat{p}$ momentum operator, $m^*$ effective mass), whereas in the case of photons $H_{kin}$ coincides with $H_{ph}$ of Ref.~\cite{mappaP}.}

Notice that in the electronic (photonic) implementation the spin-spin coupling between $e$ and the centers is of the Heisenberg ($XY$ isotropic) type. Despite this difference, the scheme is effective in both cases. We also point out that as witnessed by \eqs (\ref{Vel}) and (\ref{Vph}) there is no direct coupling between the centers, which are however indirectly coupled via the mediating scattering particle $e$.

As Hamiltonian (\ref{H}) does not contain any free-energy terms involving spin degrees of freedom, the energy spectrum of the unbound stationary states reduces to the energy-wavevector dispersion relation associated with the wire $E_k$, namely the spectrum of $H_{kin}$ ($k\!>\!0$ is the magnitude of the $e$'s wavevector).{For an electron we have $E_k\ug k^2/(2m^*)\ug m^* v_k^2/2$ ($v_k\ug k/m^*$ is the group velocity), whereas for a photon traveling along a waveguide with a linear dispersion law $E_k\ug v_{ph} k$ where $v_{ph}\ug v_k$ is the photon group velocity).} The unbound stationary states associated with a given $k\!>\!0$ can be exactly calculated through standard methods by solving the Schr\"{o}dinger equation and imposing the matching of the wavefunction at $x_i$'s ~\cite{ciccarello, mappaP, mappaNP}.  The knowledge of such states allows to quantify in terms of suitable Kraus operators \cite{nielsen, mappaP, mappaNP} how the interaction with a scattering particle $e$ later detected at $D$ changes the spin state of centers 1-3 \cite{mappaNP}. 

\section{Protocol} \label{protocol}

We consider a stream of particles $e$ to be incoming from the left. Each of them impinges
on centers 1-3 being eventually scattered off, either reflected or transmitted. We assume that the buffer time between the injection of two successive mediators is large enough to ensure that the mobile mediators interact with the centers one at a time. We call $\ket{\s}_e$ and $\ket{\g}_e$ the two orthogonal (pseudo) spin states of $e$. Each mediator $e$ is sent in the maximally mixed spin state $\rho_{e}\ug \openone_{e}/2\ug 1/2 (\ket{\s}_e\! \bra{\s}\!+\! \ket{\g}_e\! \bra{\g})$ and with wavevector $k$ such that 
\begin{eqnarray}  
k d_{12}=p \pi \label{res-cond-12} \,\,\,\,\,(p=1,2,...)\label{res-conds12},\\
k d_{23}=q \pi \label{res-cond-23} \,\,\,\,\,(q=1,2,...)\label{res-conds23},
\end{eqnarray}
which we will occasionally refer to as resonance conditions.
As each mobile particle reaches the interaction region $x_1\!<x\!<\!x_3$ multiple scattering and spin-flipping involving $e$ and centers 1-3 take place, with $e$ eventually scattered off, either reflected or transmitted. We label with $|m_i=-s,-s+1,...,s\rangle_i$ the $2 s\!+\!1$ eigenstates of $\hat{S}_{iz}$ ($i\ug1,2,3$). Our protocol teleports an arbitrary pure quantum state $\ket{\phi}=\sum_{m\ug-s,s}\alpha_m\ket{m} $ ($\sum_{m}|\alpha_m|^2\ug1$) from center 1, Alice, to center 3, Bob (see \fig 1). It is made out of the following three essential steps:
\begin{description}
  \item[(a)] Centers 2 and 3 are prepared in the product state $\ket{s,-s}_{23}$. The overall initial state of 1-3 is thus $|\chi_a\rangle \ug \ket{\phi}_1\ket{s,-s}_{23}$.
  \item[(b)] We set $J_{1}=0$ and $J_2 \ug J_3\ug J_b$ and send $n_{23}$ particles ($e$'s). We detect each particle at $D$.
  \item[(c)] We set $J_{3}=0$ and $J_1 \ug J_2\ug J_c$ and send $n_{12}$ particles ($e$'s). We detect each particle  at $D$.
\end{description}
Provided that $n_{12}$ and $n_{23}$ are large enough, the above process projects center 1 onto $\rho_{1}\ug \openone_{1}/(2s\!+\!1)$ and center 3 onto $\rho_{3}\ug  \ket{\phi}_3\! \bra{\phi}$ ($\openone_{i}\ug\sum_{m_i\ug-s,s}\ket{m_i}_i\! \bra{m_i}$ is the identity matrix of the $i$-th center). Some remarks are in order. First, there are no strict constraints on $n_{12}$ and $n_{23}$. Moreover, if the involved coupling strengths are large enough even $n_{12}\ug n_{23}\ug 1$ can suffice to accomplish teleportation (later we will explain the mechanism behind). We also point out that after step (a) \emph{no action} over the spin degrees of freedom of either the centers or the mediators is required to accomplish teleportation.
As each $e$ may in general be reflected back without reaching $D$, steps (b) and (c) are carried out probabilistically. Hence, the scheme has an associated success probability $P$. This is the overall conditional probability to detect at $D$ $n_{23}$ mobile mediators during step (b) and $n_{12}$ during step (c). Although, as mentioned, $n_{12}$ and $n_{23}$ need to be large enough and the mediators are sent one at a time, we will show that as $n_{12}$ and $n_{23}$ grow up $P$ asymptotically converges to a finite value.{We now comment on the required setting of the coupling strengths. While in general the protocol allows to have  $J_b\!\neq\! J_c$, clearly such situation would in particular demand some implicit ability to tune the coupling strength $J_2$, which appears an unnatural requirement given the limited-control scenario within which our proposal is intended to take its place. However, one may envisage a more realistic situation where the three coupling strengths $J_1$, $J_2$ and $J_3$ are all equal, even approximately (later we show the resilience of the scheme to static disorder). In such a case (corresponding to $J_b\simeq J_c$) to carry out the scheme we simply demand the ability to switch off the $e$-1 and $e$-3 couplings during steps (b) and (c), respectively (later in \secn \ref{resilience} we discuss how this can be achieved in both the implementations).}

As a further figure of merit to quantify the performance of our scheme we use the fidelity of the state of center 3 with respect to the state to be teleported $F$. Later, in \secn \ref{mechanism}, we will illustrate the approach we used to calculate $F$ and $P$.

In \figs 2(a) and (b) we address the electronic set-up [\cfr \eq(\ref{Vel})], set $J_b\ug J_c\ug1.5 v_k$, $s\ug1/2$ (spin-1/2 centers) and plot the average fidelity and probability $F$ and $P$, respectively, as functions of $n_{12}$ and $n_{23}$ (we average over all possible single-qubit pure states; $v_k$ is the group velocity of $e$). When both $n_{12}$ and $n_{23}$ are large enough, teleportation is achieved with unit fidelity ($F\!\simeq \!1$) with success probability approaching $P\ug1/8${(later on we show a strategy to double this probability).}
\begin{figure}
\centerline{\includegraphics[width=0.24\textwidth]{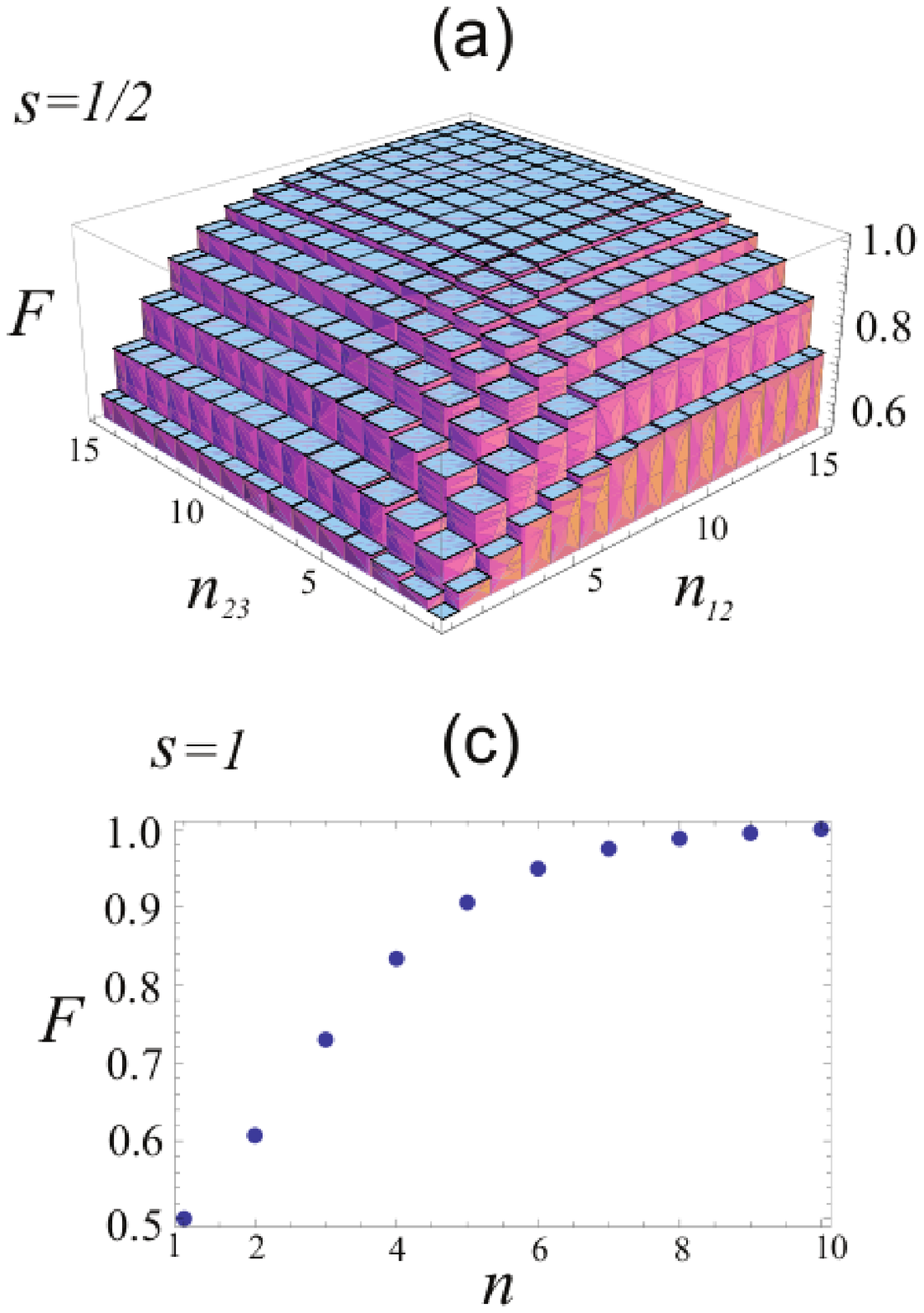}\hskip0.3cm\includegraphics[width=0.24\textwidth]{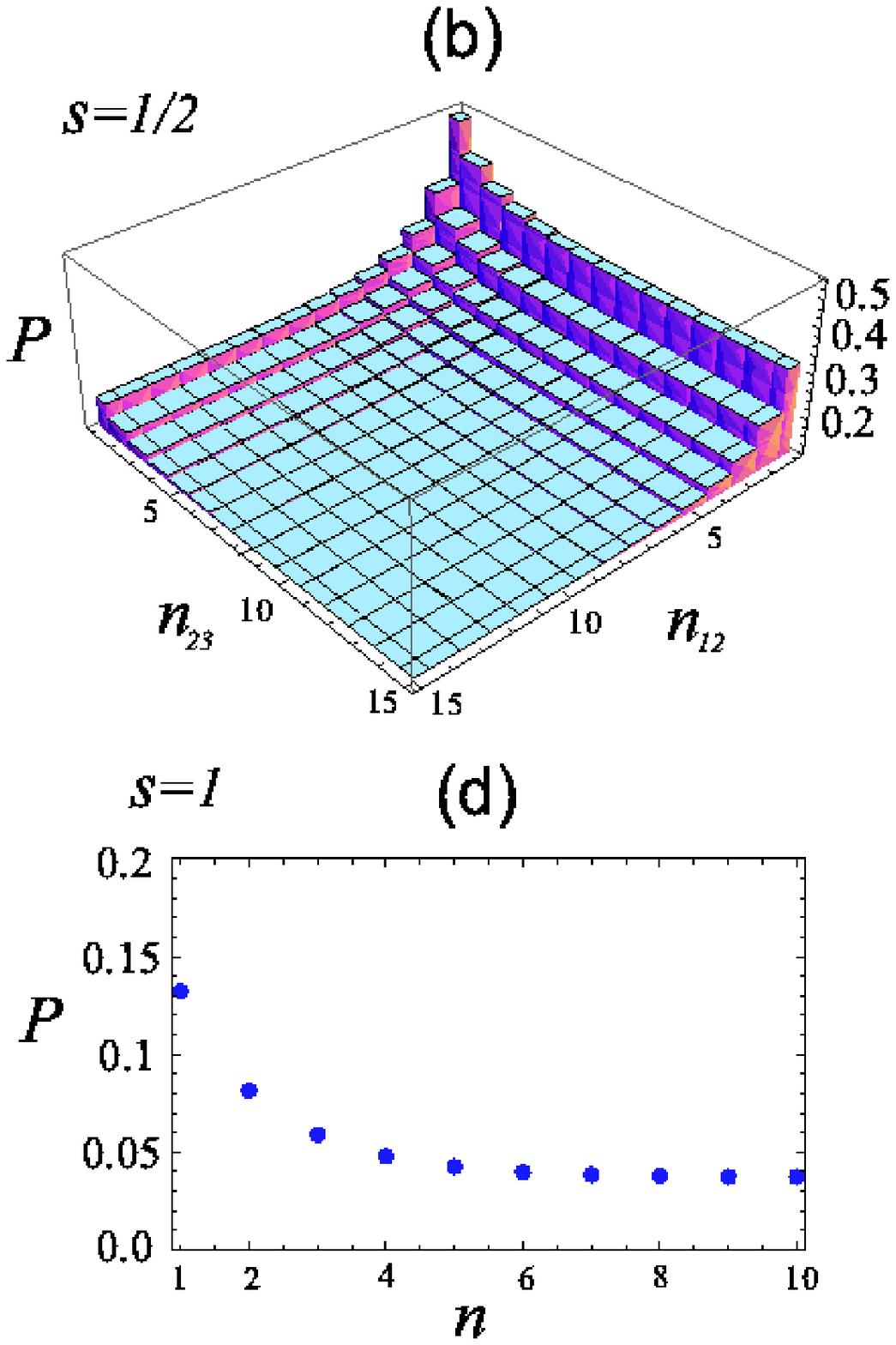}}
\caption{(Color online) Electronic setup, $s\ug1/2$: average fidelity $F$ (a) and average success probability $P$ (b)
vs. $n_{12}$ and $n_{23}$ for $J_b\ug J_c\ug1.5 v_k$. Electronic setup, $s\ug1$
average fidelity $F$ (c) and average success probability $P$ (d) vs. $n\ug n_{12}\ug n_{23}$ for $J_b\ug J_c\ug5 v_k$.  In each of these plots we have set conditions (\ref {res-cond-12})-(\ref {res-cond-23}).} \label{Fig2}
\end{figure}
The protocol's performance is stable against the number of injected mediators: The only requirement is to send and detect at $D$ a large enough number of mediators. For the ratio $J/v_k\ug1.5$ set in \fig 2 we need $n_{23}\ge 6$ and $n_{12}\ge 6$ in order to have $F\ge0.95$. Such minimum number of mediators can however be lowered by setting a larger coupling strength $J/v_k$.
\begin{figure}
\centerline{\includegraphics[width=0.25\textwidth]{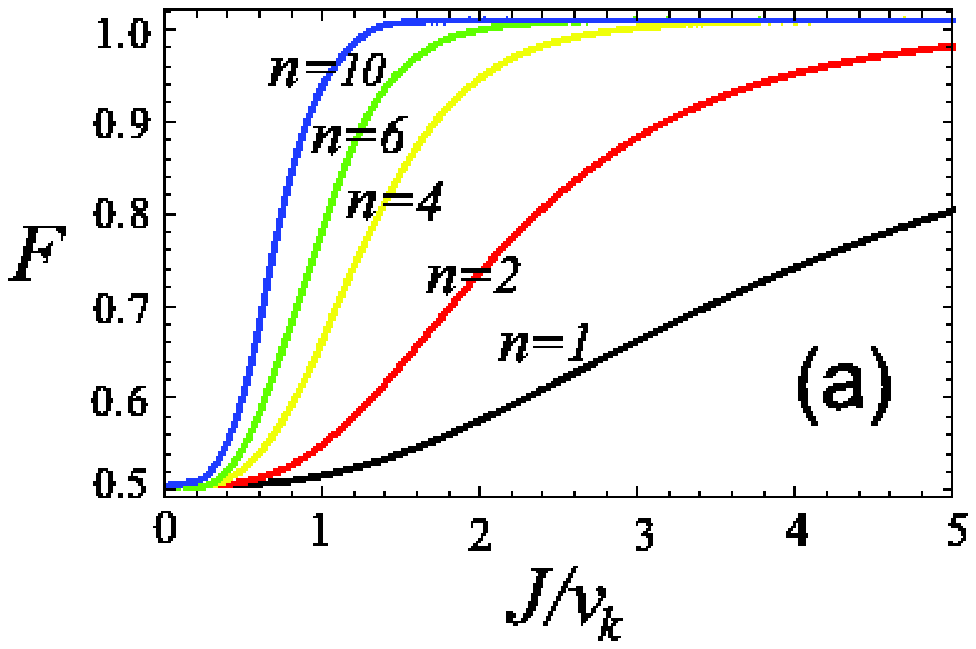}\hskip0.0cm\includegraphics[width=0.25\textwidth]{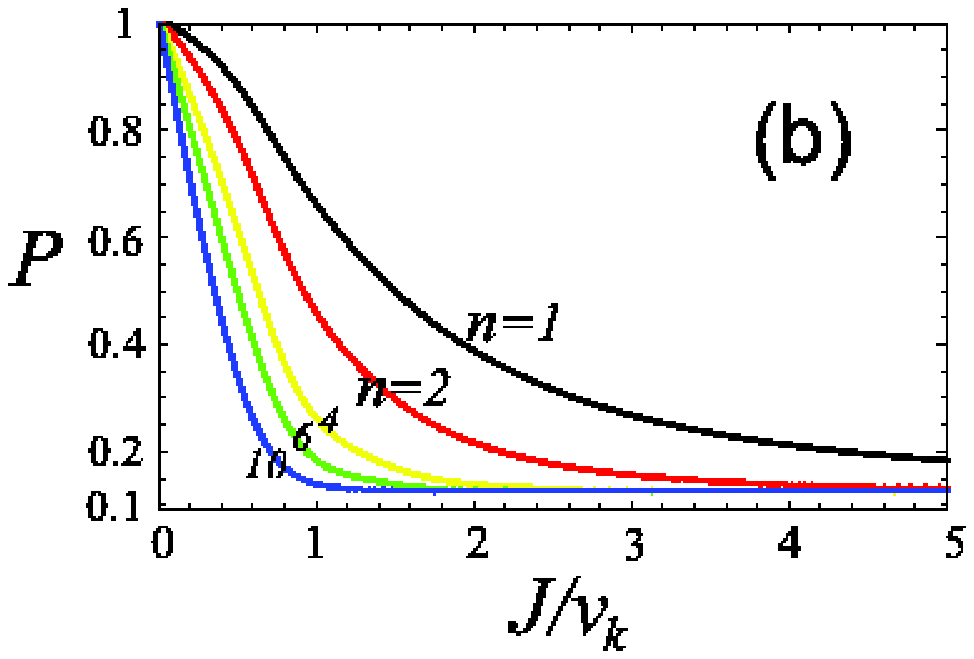}}
\caption{(Color online) Electronic setup, $s\ug1/2$: Average fidelity $F$ (a) and average success probability $P$ (b)
vs.  $J/v_k$ for $n\ug$1 (black), $n\ug2$ (red), $n\ug4$ (yellow), $n\ug6$ (green) and $n\ug10$ (blue). We have set $s=1/2$, $n\ug n_{12}\ug n_{23}$, $J\ug J_{b}\ug J_{c}$ and conditions (\ref {res-cond-12})-(\ref {res-cond-23}).} \label{Fig4}
\end{figure}
This is shown in \fig 3 where for the case $s\ug1/2$ we take $ n_{12}\ug n_{23}\ug n$, $J_b\ug J_c\ug J$ and plot $F$ and $P$ against $n$ and $J/v_k$. First, $F$ and $P$ asymptotically converge to 1 and 1/8, respectively, regardless of the coupling strength (this affects only the convergence rate).  As $J/v_k$ grows up less mediators are needed to achieve teleportation down to the minimum number $n_{12}\ug n_{23}\ug 1$ as we have checked (not shown in the figure). Likewise the dependence on $n_{23}$ and $n_{12}$, notice that no tight constraints over the coupling strengths are required. Indeed, \emph{any} coupling strength allows for teleportation provided that a large enough number of particles $e$ are sent and detected at $D$.

A further advantage of the protocol is that its working principle is independent of the center spin number $s$. Hence, it allows for teleportation of quantum states of \emph{qudits}, \ie particles whose spin space dimension is $d\ug (2s\!+\!1)$. This is illustrated in \figs 2(c) and (d) in the case $s\ug 1$ for the electronic setup. We set $n\ug n_{12}\ug n_{23}$, $J/v_{k}\ug5$ and plot the average fidelity and probability $F$ and $P$, respectively, against $n$. While $F\!\rightarrow\!1$ as $n$ grows up, $P$ converges to 1/27. Once again, notice the mild dependence on $n$, provided that the number of injected particles is large enough. Although not shown here, we have checked that a dependence on $J/v_{k}$ qualitatively analogous to that obtained for $s\ug1/2$ is exhibited in this case as well.

Finally, although the plots shown in this section refer to the electronic setup, we have checked that analogous behaviors are exhibited in the case of the photonic setup [\cfr \eqs (\ref{H}) and (\ref{Vph})].

\section{Approach and working principle of the scheme} \label{mechanism}

As anticipated in \secn \ref{setup}, all of the spin-dependent reflection and transmission probability amplitudes can be derived by imposing boundary conditions on the wavefunction and its derivative at $x\ug x_1$, $x_2$ and $x_3$ (the procedure is the natural extension to the case of three centers of the method adopted in Refs.~\cite{ciccarello,mappaP,mappaNP}). The knowledge of such coefficients allows to define suitable Kraus operators $\hat{R}_{m_e}^{m'_e}$ ($\hat{T}_{m_e}^{m'_e}$) that describe how the spin state of the centers is changed upon scattering by a particle $e$ injected in the spin state $\ket{m'_e\ug\,\s,\g}_e$ and reflected (transmitted) in the state $\ket{m_e\ug\,\s,\g}_e$ \cite{mappaP,mappaNP}. For a given initial spin state of $e$ $|m'_e=\s,\g\rangle_e$ such Kraus operators fulfill
\begin{equation} \label{closure}
\sum_{m_e\ug \s,\g} \left(\hat{R}_{m_e}^{m'_e\,\dag}\hat{R}^{m'_e}_{m_e}+\hat{T}_{m_e}^{m'_e\dag}\hat{T}^{m'_e}_{m_e} \right)=\openone_{123}\,\,,
\end{equation}
where $\openone_{123}\ug\openone_{1}\openone_{2}\openone_{3}$ ($\openone_{i}$ is the identity operator in the spin space of center $i$). Although not explicitly shown by our notation, $\hat{R}_{m_e}^{m'_e}$ and $\hat{T}_{m_e}^{m'_e}$ implicitly depend on the coupling strengths $J_i$'s, the wavevector $k$ and the distances $d_{12}$, $d_{23}$ \cite{pattern}.

The interaction of centers 1-3  with one particle $e$ injected in the spin state $\rho_e\ug\openone_{e}/2$ and wavevector matching \eqs (\ref{res-conds12}) and (\ref{res-conds23}), which is successfully detected at $D$ once it has been scattered off (see \fig 1) changes a centers' initial state $\rho_0$ into \cite{nielsen, mappaNP}
\begin{equation}\label{map}
\varrho^{(1)}=\mathcal{E}(\rho_0)= \left( 1/2 \sum_{m_e'\ug\s,\g}\sum_{m_e\ug\s,\g}\hat{T}^{m_e'}_{m_e}\rho_0 \hat{T}_{m_e}^{m_e'\,\dag}\right) /P(\rho_0)\,\,,
\end{equation}
where
\begin{eqnarray}
	P(\rho_0)=\tr \left[ 1/2 \sum_{m_e'\ug\s,\g}\sum_{m_e\ug\s,\g}\hat{T}^{m_e'}_{m_e}\rho_0\hat{T}_{m_e}^{m_e'\,\dag} \right ] \,\,\,\label{pt}
\end{eqnarray} 
is the associated probability. Eq. (\ref{map}) defines a quantum map in the spin space of 1-3 in terms of Kraus operators $\hat{T}^{m_e'}_{m_e}$'s \cite{mappaNP}.
When $n\!\ge\!1$ mobile particles are successively detected at $D$, the final spin state of the centers is obtained by $n$-time application of the map (\ref{map}) as 
\begin{equation}\label{map-n-times}
\varrho^{(n)}=\mathcal{E}^{(n)}(\rho_0)=\underbrace{\mathcal{E}\,[\mathcal{E}\,[ \cdot \cdot \, [\mathcal{E}}_n\,(\rho_0)]\,\,
\end{equation}
with overall conditional probability
\begin{equation} \label{prob-n-times}
P^{(n)}(\rho_0)=\prod_{\mu=1,n}
P(\varrho^{(\mu-1)})
\end{equation} 
with $\varrho^{(0)}\ug \rho_0$.

More specifically, our protocol consists of iterated applications of two quantum maps $\mathcal{E}_{23}$ and $\mathcal{E}_{12}$ during steps (b) and (c), respectively  (\cfr \secn \ref{protocol}). Map $\mathcal{E}_{23}$ (map $\mathcal{E}_{12}$) is map (\ref{map}) with the setting $J_1\ug0$, $J_2\ug J_3\ug J_b$ ($J_3\ug0$, $J_1\ug J_2\ug J_c$) along with \eqs (\ref{res-conds12}) and (\ref{res-conds23}). The initial spin state of 1-3 is $|\chi_a\rangle\langle\chi_a|\ug|\phi\rangle_1\langle\phi|\,|s,-s\rangle_{23}\langle s,-s|$ [step (a) in \secn \ref{protocol}]. Hence, at the end of step (c) (\cfr \secn \ref{protocol}) the final centers' state is
\begin{equation}\label{final-state}
\rho_f=\mathcal{E}_{12}^{(n_{12})}[\mathcal{E}_{23}^{(n_{23})}(|\chi_a\rangle\langle\chi_a|)]=\mathcal{E}_{12}^{(n_{12})}\circ\,\mathcal{E}_{23}^{(n_{23})}(|\chi_a\rangle\langle\chi_a|)\,\,,
\end{equation}
with success probability
\begin{equation} \label{success-P}
P_{\phi}=\prod_{\nu=1,n_{12}}
P_{12}(\varsigma^{(\nu-1)})\prod_{\mu=1,n_{23}}
P_{23}(\varrho^{(\mu-1)})\,\,
\end{equation} 
where $\varrho^{(0)}\ug |\chi_a\rangle\langle\chi_a|$, $\varrho^{(\mu)}\ug \mathcal{E}_{23}^{(\mu)}(|\chi_a\rangle\langle\chi_a|)$, $\varsigma^{(0)}\ug \varrho^{(n_{23})}$ and $\varsigma^{(\nu)}\ug\mathcal{E}_{12}^{(\nu)}(\varrho^{(n_{23}})$.

The fidelity of the final state of center 3 with respect to the state to be teleported $|\phi\rangle$ is computed as
\begin{equation} \label{fidelity}
F_{\phi}=\langle \phi|_{3}\tr_{12}\left[\mathcal{E}_{12}^{(n_{12})}\circ\,\mathcal{E}_{23}^{(n_{23})}(\rho_0) \right]|\phi\rangle_{3}\,\,,
\end{equation}
where  $\tr_{12}$ stands for the partial trace over 1 and 2.

Average fidelity and probability $F$ and $P$ in \figs 1 and 2 (\secn \ref{protocol}) and in \figs 3 and 4 (\secn \ref{resilience}) are obtained by averaging over all possible pure states \cite{average} for a given spin number $s$. For $s\ug1/2$, this family of states has the form $|\phi\rangle\ug \!\cos \vartheta/2 |\!\s\!\rangle\!+\!e^{i \varphi}\sin \vartheta/2 |\!\g \! \rangle$ ($\vartheta\ug[0,\pi]$, $\varphi\ug[0,2\pi]$). For $s\ug1$,  we use the four-angle parametrization of a pure state belonging to the generalization of the Poincar\'e sphere to the case of a qutrit  \cite{poincare}.

We call $|\Psi_s^{-}\rangle_{jl}\ug \sum_{m\ug-s,s}(-1)^\eta (2s\!+\!1)^{-1/2}|m,-m\rangle_{jl}$ the spin-$s$ singlet state of centers $j\!-\!l$ [$j$,$l\ug 1,2,3$, $\eta\ug m\!+\!1/2 \,\,(\eta\ug m)$ for $m$ semi-integer (integer)]. For $s\ug1/2$ we retrieve the standard Bell state $\ket{\Psi^-}_{jl}$. Moreover, we define a two-center interaction operators $\hat{V}_{j\neq l}$ ($j,l\ug1,2,3$), whose form for the electronic setup [\cfr \eq (\ref{Vel})] is
\begin{equation}
\hat{V}_{j\neq l}=J\,\hat{\mbox{\boldmath$\sigma$}}\cdot \left[\hat{\mathbf{S}}_j\,\delta(x-x_j)+\hat{\mathbf{S}}_l\,\delta(x-x_l)\right]\,\,,
\end{equation}
whereas for the photonic setup [\cfr \eq (\ref{Vph})] we have  
\begin{equation}
\hat{V}_{j\neq l}\ug J\,\sum_{\beta\ug x,y}\hat{\sigma}_{\beta} \left[\hat{S}_{j\beta} \,\delta(x-x_j)+ \hat{S}_{l\beta} \,\delta(x-x_l)\right]\,\,.
\end{equation}

Hence, at steps (b) and (c) the interaction term in \eq (\ref{H}) respectively reduces to $\hat{V}_{23}$ (with $J\ug J_b$) and $\hat{V}_{12}$ (with $J\ug J_c$).

It can be straightforwardly demonstrated \cite{ciccarello,mappaP,mappaNP} that if conditions (\ref{res-conds12})-(\ref{res-conds23}) are set then $\hat{V}_{23}|\Psi_s^{-}\rangle_{23}\ug \hat{V}_{12}|\Psi_s^{-}\rangle_{12}\ug 0$, \ie the singlet $|\Psi_s^{-}\rangle_{23}$ ($|\Psi_s^{-}\rangle_{12}$) gives rise to an effective quenching of the interaction between $e$ and the involved centers at stage (b) [stage (c)]. Remarkably, this occurs \emph{regardless} of the spin state of the mobile particle $e$, which can in particular be sent even in the maximally mixed state $\openone_e/2$ \cite{ciccarello,mappaNP} as we have assumed so far. Such effective quenchings necessary entail that \cite{nota-hilbert}
\begin{eqnarray}
\mathcal{E}_{23}\left(|\Psi^-_s\rangle_{23}\langle\Psi^-_s|\right)&=&|\Psi^-_s\rangle_{23}\langle\Psi^-_s|\,\,,\\
\mathcal{E}_{12}\left(|\Psi^-_s\rangle_{12}\langle\Psi^-_s|  \right)&=&|\Psi^-_s\rangle_{12}\langle\Psi^-_s|\,\,
\end{eqnarray}
with probabilities 
$P_{23}(|\Psi^-_s\rangle_{23}\langle\Psi^-_s|)\ug P_{12}(|\Psi^-_s\rangle_{12}\langle\Psi^-_s|)\ug 1$. In other words, such singlets are fixed points of the respective quantum maps. As there are no other states belonging to the spin space of centers 2-3 (1-2) with the same features \cite{mappaNP} we conclude that the successive detections of a large enough number of mobile particles at $D$ achieves an effective projective measurement of the singlet $|\Psi^-_s\rangle_{23}\langle\Psi^-_s| (|\Psi^-_s\rangle_{12}\langle\Psi^-_s|)$. Such behaviour is harnessed to establish maximum entanglement between centers 2 and 3 at step (b) (see \secn  \ref{protocol}) starting from the initial state $|\chi_a\rangle \ug \ket{\phi}_1\ket{s,-s}_{23}$ prepared at step (a). Provided that $n_{23}$ is large enough, such product state is projected onto  $|\chi_b\rangle\ug\ket{\phi}_1\ket{\Psi^-_s}_{23}$ with probability $P_{23}\ug |\langle \Psi^-_s|s,-s\rangle|^2\ug(2s\!+\!1)^{-1 }$ [recall that for any $m\ug-s,...,s$ we have $\langle \Psi^-_s|m,-m\rangle\ug(-1)^m (2s\!+\!1)^{-1/2 }$]. In the final stage (c), provided that $n_{12}$ is large enough effective projection of  $|\chi_b\rangle$ onto $|\Psi^-_s\rangle_{12}$ is accomplished, which yields
\begin{equation} \label{projection-c}
_{12}\langle \Psi^-_s|\chi_b\rangle=\frac{1}{2s+1}\,|\phi\rangle_{3}\,\,,
\end{equation}
up to an irrelevant phase-factor. The initial state $|\phi\rangle$ of center 1 is therefore teleported to center 3. The success probability associated with the latter projection is thus $P_{12}\ug1/(2s+1)^2$ so that the overall success probability is $P\ug P_{12} P_{23}\ug1/(2s+1)^3$ [for $s\ug1/2$ ($s\ug1$) we obtain $P\ug1/8$ ($P\ug1/27$)]. Notice that  our proposal can in fact be effectively mapped into the seminal protocol by Bennett \emph{et al.} \cite{bennett}.

The reason why the number of required mediators to make teleportation effective decreases as the coupling strength grows up is analogous to the mechanism behind entanglement generation via the map (\ref{map}) \cite{mappaNP,yuasa}. The larger the coupling strength the more efficient is singlet-state discrimination via detection of transmitted particles $e$'s. This is because for any two-center spin state other than the singlet the transmittivity of particle $e$ decreases as the coupling strength is made larger. Hence, less transmitted electrons need to be recorded at $D$ in order to discriminate $|\Psi_{s}^-\rangle$ \cite{ciccarello, mappaNP,yuasa}.

We stress that, although probabilistic, our protocol is of a conclusive nature and basically implements the teleportation task according to the very seminal protocol by Bennett \emph{et al.} (aside from the ability to discriminate only one maximally entangled state). Therefore, at the end of stage (b), if successful, centers 2-3 are in a maximally entangled pure state, which provides the essential resource for teleportation. At the end of stage (c), if successful, centers 1-2 are projected onto a maximally entangled pure state to accomplish quantum information transfer. In other words, teleportation is in fact achieved through effective preparation and projection onto pure entangled states.

\section{Implementation issues} \label{resilience}

As discussed in the previous Section, the mechanism behind the scheme harnesses the singlet-induced spin-spin quenching, \ie $\hat{V}_{12}|\Psi_s^{-}\rangle_{12}\ug \hat{V}_{23}|\Psi_s^{-}\rangle_{23}\ug 0$. Such behavior does not take place for any pattern of the parameters entering the dynamics. Indeed, in order for it to occur we require that  conditions (\ref{res-cond-12})-(\ref{res-cond-23}) are fulfilled and that $J_2\ug J_3$ ($J_1\ug J_2$) at step (b) [at step (c)]. These constraints ensure that during stage (b) [stage (c)] the coupling of centers 2 and 3 (1 and 2) to each particle $e$ occurs \emph{symmetrically}, which entails singlet-state induced quenching of the scattering potential $V$ in \eq (\ref{H}), as it can be straightforwardly proved \cite{ciccarello, mappaP, mappaNP}. From the experimental perspective, a natural question to ask is: how resilient is the teleportation scheme against an imperfect setting of such a symmetrical situation? Here, we thus scrutiny the robustness against an imperfect setting of conditions (\ref{res-cond-12})-(\ref{res-cond-23}) as well as static disorder affecting the coupling strengths. 

We model an imperfect setting of conditions (\ref{res-cond-12})-(\ref{res-cond-23}) by assuming that the wavevector of an injected mediator $k$ is selected according to a Gaussian distribution around a carrier wavevector $k_0$ with standard deviation $\Delta k$. Resonance-conditions in \eqs  (\ref{res-cond-12})-(\ref{res-cond-23}) are thus exactly fulfilled only by the carrier wavevector $k\ug k_0$ (\ie the ideal case addressed in \secns \ref{protocol} and \ref{mechanism}  is recovered by setting $\Delta k\ug 0$). As a paradigmatic instance, we consider the photonic setup [see \secn \ref{setup}, \eqs (\ref{H}) and (\ref{Vph})] and set $s\ug1/2$, $n_{12}\ug n_{23}\ug 8$ (in such a case the mediator's velocity is independent of $k$ according to $v_k\ug v_{ph}$). In the ideal situation (see \secn \ref{protocol}) by taking $J_b\ug J_c\ug 2 v_{ph}$ and resonance-conditions (\ref{res-cond-12})-(\ref{res-cond-23}) we obtain the average fidelity and probability respectively as $F\!\simeq\! 0.96$ and $P\!\simeq\! 0.13$. We call $J_{3b}$ and $J_{1c}$ the values of $J_3$ and $J_1$ during steps (b) and (c), respectively [during step (b) $J_1\ug 0$, whereas during step (c) $J_3\ug0$, see \secn \ref{protocol}]. In \fig 5 we take $J_2\ug  2 v_{ph}$ and plot $F$ and $P$ against $\Delta k$ for the symmetrical pattern $J_{3b}\ug J_{1c}\ug J_2$ and the asymmetrical one $J_{3b}\ug 0.9\%J_{2}$ and $J_{1c}\ug 1.01 \%J_2$ (\ie for deviations of $J_{3b}$ and $J_{2c}$ of, respectively, $10\%$ and 1$\%$ from their ideal values).
\begin{figure}
\centerline{\includegraphics[width=0.50\textwidth]{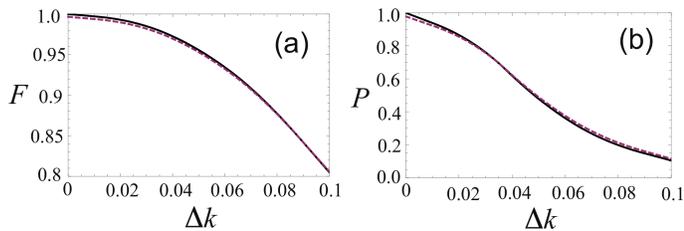}} 
\caption{Photonic setup, $s\ug1/2$: average fidelity $F$ (a) and average success probability $P$ (b) against $\Delta k$ (in units of $k_0$) for $J_{3b}\ug J_{1c}\ug J_2\ug2 v_{ph}$ (symmetrical couplings, solid line) and $J_2\ug 2 v_{ph}$, $J_{3b}\ug 0.9\%J_{2}$ and $J_{1c}\ug 1.01 \%J_2$ (asymmetrical couplings, dashed line). Only the carrier wavevector $k_0$ of each mobile particle fulfills conditions (\ref {res-cond-12})-(\ref {res-cond-23}). $F$ and $P$ are normalized to their values for $\Delta k\ug 0$ and the symmetrical pattern of coupling strengths $J_{3b}\ug J_{1c}\ug J_2\ug  2 v_{ph}$. \label{Fig4}}
\end{figure}
First, notice that the scheme's performances are only slightly affected in the case of the asymmetrical pattern of coupling strengths: while the average fidelity is basically insensitive, the maximum deviation of probability is $\!\simeq\!3\%$ for $\Delta k/k_0$ up to  $\!\simeq\!5\%$ and $\!\simeq\!10\%$ for $\Delta k/k_0$ between $\!\simeq\!5\%$ and $\!\simeq\!10\%$. As for the dependence on $\Delta k$, both $F$ and $P$ monotonically decrease as the wavevector uncertainty grows up. For values of $\Delta k/k_0$ up to $0.1$, the maximum decrease of $F$ ($P$) is $\simeq\!20\%$ ($\simeq\!90\%$). However, for $\Delta k/k_0$ up to $\simeq\! 3\%$ the maximum decrease of $F$ is lower than $5\%$, whereas that of $P$ is $\simeq\!30\%$. Overall, the protocol therefore exhibits quite a striking resilience against static disorder in the pattern of coupling strengths and a reasonable tolerance against an imperfect matching of resonance conditions (\ref {res-cond-12})-(\ref {res-cond-23}). The outcomes of this analysis are in line with similar studies carried out to test the resilience of some scattering-based entanglement generation schemes  \cite{resil-refs} and provide further evidence of the low level of control required by scattering-based methods to perform QIP tasks.

We now comment on the switch-off of the coupling between the mobile particle $e$ and center 1 (center 3) required at step (b) [step (c)] (see \secn \ref{protocol}). In an electronic implementation using electrostatically-defined  single-electron QDs \cite{single-electron} to embody the centers the Heisenberg-type coupling in \eq (\ref{Vel}) arises from the exchange interaction between the bound and propagating electrons. In such a case, the QD orbitals, and hence the exchange interaction, can be controlled, and thus switched off, by tuning gate voltages. Alternatively, as it suffices the dots lie close to the wire along which each particle $e$ can travel (without the need for being embedded within it) one can envisage the setup depicted in \fig 5. It is made out of three centers, two wires and two detectors.The centers are arranged on the $z\ug 0$ plane according to an L geometry. A wire on the $z\ug 0$ plane lies close to centers 1 and 2 but far from center 3, whereas another wire lies on a plane $z\ug z_0\neq0$ close to centers 2 and 3 but far from 1. One detector is placed at the end of each wire. The advantage of this setup is that if an electron is injected in the $z\ug 0$ ($z\ug z_0$) wire then it interacts only with centers 1 and 2 (2 and 3). The teleportation protocol would work analogously to the one for the setup in \fig 1, the only modification being that at stage (b) [stage (c)] the injected electrons would flow through the wire on the $z\ug z_0$ ($z\ug 0$) plane. 

\begin{figure}
\centerline{\includegraphics[width=0.35\textwidth]{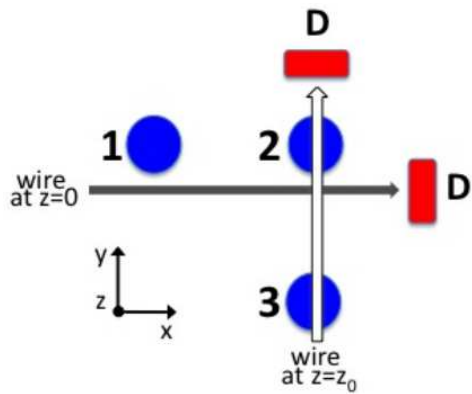}} 
\caption{(Color online) Alternative electronic setting. All of the centers and one of the wires lie on the $z\ug 0$ plane, whereas the other wire lie on a $z\ug z_0$ plane ($z_0\neq 0$). A path detector is placed at the end of each wire.}
\end{figure}

In a photonic implementation using multi-level QDs to embody the centers \cite{mappaP,mappaNP} the switch-off of the effective $XY$ coupling between $e$ and one of the centers in \eq (\ref{Vph}) can be obtained by setting the photon-QD interaction far off-resonance, \eg via a Zeeman shift of the involved QD levels.  

As for the effect of unavoidable decoherence processes affecting the centers, although its detailed analysis is beyond the scopes of this work we expect our protocol to be quite robust against this. First, as discussed in \secn \ref{mechanism} our procedure in fact relies upon effective projective measurements of two-center singlet states, which motivates such expectation in the case that the distance between the centers is not much larger than the environmental correlation length. Second, in the case of not too weak coupling strengths the scheme can be effective with only a few number of injected mobile particles (see \fig 3 and related discussion) (for issues related to decoherence see also the discussion in Ref.~\cite{mappaNP}).

\bigskip

\section{Conclusions} \label{conclusions}

In this work we have presented a general setting where teleportation of an unknown qudit state between two remote scattering centers is achieved solely via repeated injections of unpolarized mobile particles undergoing multiple scattering along with path detection. Teleportation is carried out without taking any action over the quantum internal state of both the centers and the flying particles. No strict constraints over both coupling strengths and interaction times are demanded. We only require each mediator to have enough time to be scattered from the centers and its wavevector to fulfill appropriate resonance conditions, whose matching, though, is quite robust against an imperfect setting and/or static disorder. Our scheme is particularly well-suited to scenarios with only very limited access to quantum memories. The mechanism behind our proposal essentially relies on resonance-induced symmetries and the conditioning effect over the system's dynamics due to path detection. An implicit requirement is that the interaction between each mobile particle and the centers occurs coherently: despite its counter-intuitiveness this suffices to make teleportation effective even though each flying particle is injected in a maximally mixed state and without any post-selection over its internal state. In the case that the mobile particles and centers are respectively embodied by photons and multi-level atoms, whose interaction occurs according to a Jaynes-Cummings-like model, the counter-intuitiveness is strengthened by the requirement that the polarization state of each injected photon \emph{needs} to be mixed to make teleportation effective \cite{nota-mappaP}. In some sense, our protocol shares some features with measurement-based models of quantum computation. Indeed, from the viewpoint of Quantum Information Theory, the harnessed resource is ultimately the entanglement between the centers established by their successive interactions with the injected mediators. Here, however, the local measurements, embodied by the{path} detection over the mobile particles, have the twofold purpose of both creating such a computational resource and processing information once it has been built up.
 
{To our knowledge, this is the first scheme where the spatial movement of quantum information is accomplished by scattering processes. We recall that we do not require either a direct cross-talk between the centers or a direct Bell-state measurement to be performed on them. }

{Furthermore, our work shows that in the case that the centers' spin number is higher than 1/2 \emph{qubit} mediators can allow for transmission of a pure quantum state between two \emph{high-dimensional} entities.}

We point out that the singlet-state extraction scheme in Ref.~\cite{mappaP}, where \emph{polarized} mobile mediators are post-selected{either reflected or transmitted in their initial spin state (\ie spin-state post-selection is carried out),} cannot be used to perform teleportation. The reason is that in such a case the singlet is not the only fixed point of the associated quantum map  \cite{mappaP, nota-mappaP}. Its iterated application is thus not equivalent to a Bell-state measurement within the \emph{whole} involved-centers' spin space [\ie the spin space of centers 2-3 at step (b) and that of 1-2 at step (a)] as it takes place for map in \eq (\ref{map}). 

As for the success probability of the present scheme, we point out that although significant ($P\ug1/8$ in the spin-1/2 case) this is lower than those achieved by the most of the probabilistic teleportation proposals appeared in the literature \cite{telep-QED, no-bell} (where typically $P\ug1/4$ or $P\ug1/2$). It should be remarked, though, that the level of control required by the present proposal is extremely low.{Furthermore, latest findings on setups analogous to those addressed here \cite{daniel} have shown that a suitable modification of schemes in Refs.~\cite{ciccarello,yuasa} can allow for deterministic singlet-state preparation. This strategy can deterministically replace steps (a) and (b) of our procedure (see \secn \ref{protocol}), \ie the entire entanglement-preparation stage \cite{nota-daniel}. In such a case, the teleportation protocol would have an associated success probability of 1/4 (see discussion related to \eq \ref{projection-c}).}

To our knowledge, this is the first proposal that uses iterated applications of a quantum map to perform QIP tasks other than generation of non-classical states. We believe this feature may be enlightening for the design of QIP protocols (even in physical scenarios different from those addressed here) that are able to work in situations of limited control, especially within the framework of computation models where quantum information is processed through measurements rather than gate-based operations. Furthermore, our scheme shows a fully scattering-based strategy to perform a quantum algorithm of pivotal importance in QIP theory \cite{tbqc}. As such, we believe that it may represent a significant milestone for future developments of a scattering-based model of quantum computation, a topic which has been the focus of recent findings in the context of quantum random walks \cite{childs}.

\bigskip
\begin{acknowledgments}

We acknowledge fruitful discussions with M.~Paternostro, G.~M.~Palma and D.~Burgarth. S.~Bose acknowledges support from the Engineering and Physical Sciences Research Council
(EPSRC) UK for an Advanced Research Fellowship, the Royal Society and the Wolfson
Foundation.

\end{acknowledgments}


\begin {thebibliography}{99}
\bibitem{bennett} C. H. Bennett, G. Brassard, C. Cr\`epeau, R. Jozsa, A. Peres, and W. K. Wootters, Phys. Rev. Lett. \textbf{70}, 1895
(1993).
\bibitem{nielsen} M. A. Nielsen and I. L. Chuang,  \textit{Quantum Computation and
Quantum Information} (Cambridge University Press, Cambridge, U. K.,
2000).
\bibitem{epr} A. Einstein, B. Podolsky, and N. Rosen, Phys. Rev. \textbf{47}, 777
(1935).
\bibitem{bowm-boschi} D. Bouwmeester, J.-W. Pan, K. Mattle, M. Eibl, H. Weinfurter, and A. Zeilinger,  Nature (London) \textbf{390}, 575
(1997); D. Boschi, S. Branca, F. De Martini, L. Hardy, and S.
Popescu, Phys. Rev. Lett. \textbf{80}, 1121 (1998); M. A. Nielsen, E. Knill, and R. Laflamme, Nature (London) \textbf{396}, 52 (1998); A. Furusawa, J. L. Sorensen, S. L. Braunstein, C. A. Fuchs, H. J. Kimble, and E. S. Polzik, Science \textbf{282}, 706 (1998); M. Riebe, H. H\"{a}ffner, C. F. Roos, W. H\"{a}nsel, J. Benhelm, G. P. T. Lancaster, T. W. K\"{o}rber, C. Becher, F. Schmidt-Kaler, D. F. V. James, and R. Blatt, Nature (London) \textbf{429}, 734 (2004); M. D. Barrett, J.
Chiaverini, T. Schaetz, J. Britton, W. M. Itano, J. D. Jost, E.
Knill, C. Langer, D. Leibfried, R. Ozeri, and D. J. Wineland, Nature
(London) \textbf{429}, 737 (2004); J. Sherson, H. Krauter, R. K. Olsson, B. Julsgaard, K. Hammerer, I. Cirac, E. S. Polzik, Nature (London) \textbf{443}, 557 (2006); Q. Zhang, A. Goebel, C. Wagenknecht, Y.-A. Chen (Chen, Yu-Ao), B. Zhao, T. Yang T, A. Mair, J. Schmiedmayer, J. W. Pan, Nat. Phys. \textbf{2}, 678 (2006).
\bibitem{tbqc} D. Gottesman, and I.L. Chuang,
Nature (London) \textbf{402},  390(1999); E. Knill, R. Laflamme, and G. J. Milburn, Nature (London) {\bf409}, 46 (2001); M. A. Nielsen, Phys. Lett. A 308, {\bf 96} (2003); DD. W. Leung, Int. J. Quantum Inf. {\bf 2}, 33 (2004).\bibitem{browne} H. J. Briegel, D. E. Browne, W. D\"ur, R. Raussendorf, M. Van den Nest,  Nat. Phys. (London) \textbf{5}, 19 (2009).
\bibitem{telep-QED} L. Davidovich, N. Zagury, M. Brune, J. M. Raimond, and S. Haroche, Phys. Rev. A \textbf{50}, R895 (1994); J. I. Cirac, and A. S. Parkins, Phys. Rev. A \textbf{50},
R4441 (1994); S. B. Zheng and G. C. Guo, Phys. Lett. A \textbf{232},
171 (1997); S. B. Zheng, Opt. Commun. \textbf{167}, 111 (1999); S.
Bose, P. L. Knight, M. B. Plenio, and V. Vedral, Phys. Rev. Lett.
\textbf{83}, 5158 (1999); S. Bandyopadhyay, Phys. Rev. A 62, 012308
(2000); S. Campbell, and M. Paternostro, arXiv:0809.3583v1 [quant-ph].  
\bibitem{no-bell} L. Vaidman, Phys. Rev. A \textbf{49}, 1473 (1994); N.G. de Almeida, R. Napolitano, and M. H. Y. Moussa, Phys. Rev A. \textbf{62}, 010101 (2000); S.-B- Zheng, Phys. Rev. A \textbf{69}, 064302 (2004); L. Ye, and G.-C. Guo, Phys. Rev. A \textbf{70}, 054303 (2004); H. J. Carmichael, and B. C. Sanders, Phys. Rev. A, \textbf{60}, 2497 (1999); Hagley \emph{et al.}, Phys. Rev. Lett. \textbf{79}, 1 (1997); W. B. Cardoso, A. T. Avelar, B. Baseia, and N. G. de Almeida Phys.
Rev. A \textbf{72}, 045802 (2005); R. W. Chhajlany and A. Wójcik, Phys. Rev. A \textbf{73}, 016302 (2006); L. Ye, and G.-C. Guo, Phys. Rev. A \textbf{73}, 016303 (2004);  M. Tumminello and F. Ciccarello, Phys. Rev. A \textbf{77}, 023825 (2008).
\bibitem{telep-ss}  L.-A. Wu and D. A. Lidar Phys. Rev. A \textbf{67}, 050303 (2003);  O. Sauret, D. Feinberg, and T. Martin, Phys. Rev. B \textbf{69}, 035332 (2004);  C. W. Beenakker and M. Kindermann, Phys. Rev. Lett. \textbf{92}, 056801 (2004); F. de Pasquale, G. Giorgi, and S. Paganelli, Phys. Rev. Lett. \textbf{93}, 120502 (2004);  M. N. Leuenberger, M. E. FlattŽ, and D. D. Awschalom, Phys. Rev. Lett. \textbf{94}, 107401 (2005);  R. L. de Visser and M. Blaauboer, Phys. Rev. Lett. \textbf{96}, 246801 (2006);  S. Y., Yechao Zhu, and . Yeo, Phys. Rev. A \textbf{77}, 062338 (2008).
\bibitem{nota-bell} Some authors explicitly adopted the terminology ``teleportation without Bell-state measurements". It should be noticed that when teleportation is achieved via local measurements, this typically occurs after that some interaction process has taken place, which means that such measurements are actually effective \emph{indirect} Bell-state measurements. This is is indeed the case of the protocol presented in this work.
\bibitem{pan-fiu} A. M. Goebel {\it et al.}, arXiv:0809.3583v1 [quant-ph];  L. Slodicka, M. Jezek, and J. Fiurasek, Phys. Rev. A \textbf{79}, 050304 (2009).
\bibitem{imps} A.T. Costa, Jr., S. Bose, and Y. Omar,
Phys. Rev. Lett.~\textbf{96}, 230501 (2006); G.L. Giorgi and F. De
Pasquale, Phys. Rev. B \textbf{74}, 153308 (2006); 
K. Yuasa and H. Nakazato, J. Phys. A: Math. Theor. \textbf{40}, 297 (2007); Y. Hida, H. Nakazato, K. Yuasa and Y. Omar, Phys. Rev. A \textbf{80}, 012310 (2009);  Y. Hida, H. Nakazato, K. Yuasa and Y. Omar, arXiv:0901.2199.
\bibitem{ciccarello} F. Ciccarello \emph{et al.}, New J. Phys. {\bf 8},
214 (2006); J. Phys. A: Math. Theor. \textbf{40}, 7993 (2007); F. Ciccarello, G. M. Palma, and M.
Zarcone, Phys. Rev. B \textbf{75}, 205415 (2007); F. Ciccarello, M. Paternostro, G. M. Palma and M. Zarcone, Phys. Rev. B \textbf{80}, 165313 (2009).
\bibitem {resil-refs} F. Ciccarello \emph{et al.}, Las.
Phys. \textbf{17}, 889 (2007); F. Ciccarello, M. Paternostro, M. S. Kim, and G. M. Palma, Int. J. Quant. Inf. {\bf 6}, 759 (2008).
\bibitem{habgood}  M. Habgood, J. H. Jefferson, G. A. D. Briggs, Phys. Rev. B~\textbf{77}, 195308 (2008); J.  Phys.: Condens.  Matter   \textbf{21} , 075503 (2009).
\bibitem{mappaP} F. Ciccarello, M. Paternostro, M. S. Kim, and G. M. Palma,~\prl~\textbf{100}, 150501 (2008).
\bibitem{mappaNP}F. Ciccarello, M. Paternostro, G. M. Palma and M. Zarcone, New J. Phys. \textbf{11}, 113053 (2009).
\bibitem{yuasa} K. Yuasa,  J. Phys. A \textbf{43}, 095304 (2010).
\bibitem{daniel} K. Yuasa, D. Burgarth, V. Giovannetti, and H. Nakazato, New J. Phys. \textbf{11}, 123027 (2009).
\bibitem{geiger} By Geiger-like detector we mean a detector able to detect the presence of an itinerant particle regardless of its spin state.
\bibitem{single-electron} J.H. Jefferson, A. Ramsak, and T. Rejec, Europhys. Lett. \textbf{75}, 764 (2006); D. Gunlycke, J. H. Jefferson, T. Rejec, A. Ramsak, D. G. Pettifor, and G. A. D. Briggs, J. Phys.: Condens. Matter \textbf{18}, S851 2006.
\bibitem{photonmodel} M. Atat\"ure, J. Dreiser, A. Badolato, A. H\"ogele, K. Karrai, and A. Imamoglu, Science {\bf 312}, 551 (2006); K. Hennessy, A. Badolato, M. Winger, D. Gerace, M. Atat\"ure, S. Gulde, S. F\"alt, E. L. Hu, and A. Imamoglu, Nature (London) {\bf 445}, 896 (2007); K.-M. C. Fu, C. Santori, P. E. Barclay, I. Aharonovich, S. Prawer, N. Meyer, A. M. Holm, and R. G. Beausoleil, arXiv:0811.0328v1 [quant-ph]; P. E. Barclay, K.-M. Fu, C. Santori, and R. G. Beausoleil, arXiv:0904.0500 [quant-ph].
\bibitem{nota-hilbert} Clearly, due to $J_1\ug0$ ($J_3\ug0$) at step (b) [step (c)] $\mathcal{E}_{23}$ ($\mathcal{E}_{12}$) effectively acts over the spin space of centers 2-3 (1-2) likewise $\hat{V}_{23}$ ($\hat{V}_{12}$).
\bibitem{pattern} We leave the pattern of such parameters unspecified until \eq (\ref{prob-n-times}).
\bibitem{average} L. Hardy and D. D. Song \pra \textbf{63}, 032304 (2001); M. D. Bowdrey, D. K. L. Oi, A. J. Short, K. Banaszek, and J. A. Jones, Phys. Lett. A \textbf{294}, 258 (2002).
\bibitem{poincare} Arvind, K. S. Mallesh, and N. Mukunda, J. Phys. A \textbf{30},
2417 (1997); C. M. Caves and G. J. Milburn, Opt. Commun. \textbf{179}, 439
(2000).
\bibitem{nota-mappaP} For polarized electrons injected in $|\!\s\rangle_{e}$ and post-selected in $|\!\s\rangle_{e}$ the centers' fixed point other than the singlet is $|\!\s,\s\rangle_{12}$. Concerning the present scheme, notice that in the photonic setup the coupling between $e$ and the centers effectively vanishes for up-polarized photons when the centers' state is $|\!\s,\s\rangle_{12}$, while it does not for down-polarized photons. Hence, in our protocol sending particles in a statistical mixture of $|\!\s\rangle_{e}$ and $|\!\g\rangle_{e}$ is crucial for its effectiveness in the photonic setting (see also Ref. \cite{mappaNP}).
\bibitem{nota-daniel} It should be pointed out that the scheme in Ref. \cite{daniel} demands an additional detector able to record back-reflected particles along with the ability to inject, depending on the detector outcomes, either on- or off-resonance mediators.
\bibitem{childs}  A. M. Childs, Phys. Rev. Lett. \textbf{102}, 180501 (2009); N. B. Lovett, S. Cooper, M. Everitt, M. Trevers, and V. Kendon, arXiv:0910.1024 [quant-ph].
\end {thebibliography}
\end{document}